\documentclass[prd,aps,showpacs,nofootinbib,nobibnotes,superscriptaddress]{revtex4}

\usepackage{graphicx}
\usepackage[figuresright]{rotating}
\usepackage{amssymb}
\usepackage{bm}
\usepackage{color}

\newcommand{\be}{\begin{equation}}
\newcommand{\ee}{\end{equation}}

\newcommand{\bea}{\begin{eqnarray}}
\newcommand{\eea}{\end{eqnarray}}
\newcommand{\bean}{\begin{eqnarray*}}
\newcommand{\eean}{\end{eqnarray*}}
\newcommand{\pkt}{\; .}
\newcommand{\kma}{\; ,}

\newcommand{\labelcaption}[2]{\caption[#1]{\label{#1}#2}}

\begin{document}

\title{Primordial Magnetic Helicity Constraints from WMAP Nine-Year Data}

\author{Tina Kahniashvili}
\email{tinatin@andrew.cmu.edu}
\affiliation{McWilliams Center for
Cosmology and Department of Physics, Carnegie Mellon University,
5000 Forbes Ave, Pittsburgh, PA 15213, USA}
\affiliation{Department of Physics, Laurentian University, Ramsey
Lake Road, Sudbury, ON P3E 2C,Canada}
\affiliation{Abastumani Astrophysical Observatory, Ilia State University,
3--5 Cholokashvili St., 0194 Tbilisi, Georgia}

\author{Yurii Maravin}
\email{maravin@phys.ksu.edu}
\affiliation{Department of Physics, Kansas State University,
116 Cardwell Hall, Manhattan, KS 66506, USA}
\affiliation{Abastumani Astrophysical Observatory, Ilia State University,
3--5 Cholokashvili St., 0194 Tbilisi, Georgia}

\author{George Lavrelashvili}
\email{lavrela@itp.unibe.ch}
\affiliation{Department of Theoretical Physics, A.~Razmadze Mathematical Institute,
I.~Javakhishvili Tbilisi State University, 0177 Tbilisi, Georgia}
\affiliation{Max Planck Institute for Gravitational Physics,
 Albert Einstein Institute, 14476 Potsdam, Germany}

\author{Arthur Kosowsky}
\email{kosowsky@pitt.edu}
\affiliation{Department of Physics and Astronomy, University of Pittsburgh,
3941 O'Hara Street, Pittsburgh, PA 15260 USA}
\affiliation{Pittsburgh Particle Physics, Astrophysics, and Cosmology Center (Pitt-PACC), Pittsburgh, PA 15260 USA}

\date{\today}

\begin{abstract}
If a primordial magnetic field in the universe has non-zero helicity, the violation of parity
symmetry results in non-zero correlations between cosmic microwave background temperature
and B-mode polarization. In this paper we derive approximations to the relevant microwave
background power spectra arising from a helical magnetic field. Using the cross-power spectrum
between temperature and B-mode polarization
from the WMAP nine-year data, we set a 95\% confidence level upper limit on
the helicity amplitude
to be 10 nG$^2$ Gpc for helicity spectral index  $n_H = -1.9$,
for a cosmological magnetic field with effective field
strength of 3 nG and a power-law index $n_B = -2.9$ near the scale-invariant value.
Future microwave background polarization maps with greater sensitivity will
be able to detect the helicity of an  inflationary magnetic field well
below the maximum value allowed by microwave background constraints on the
magnetic field amplitude.
\end{abstract}

\maketitle

\section{Introduction} \label{sec:intro}

A challenging question of modern astrophysics is the origin of observed magnetic fields
in galaxies and clusters \cite{Widrow}. Generally, fields observed today began as small
seed fields and then were amplified via either adiabatic compression or through turbulent plasma
dynamics. One mechanism for seed field generation is generic plasma instabilities and
vorticity perturbations \cite{Kulsrud:2007an}. In this causal model, the correlation length of the resulting
fields is limited by the horizon, which generically corresponds to comoving galaxy scales.
A second possibility is larger seed fields generated during inflation spanning a wide range
of correlation lengths up to the horizon today, and amplified through the process of
cosmological structure growth~\cite{Kandus:2010nw,Durrer:2013pga}.

The evolution and amplification of a primordial seed field is strongly influenced by the
helicity, or local handedness, of the seed field. Magnetic helicity is a manifestation of parity symmetry violation.
While the level of parity violation observed in fundamental physical interactions is
small, parity violation is widespread in various astrophysical systems with significant magnetic
dynamics, such as one-sided jets from active galactic nuclei
and helical magnetic fields in the solar magnetosphere~\cite{Brandenburg:2004jv}.
A seed field with helicity is restructured at large scales by plasma turbulence:
the decay of the magnetic field leads to an increase in the relative magnetic helicity until the helicity saturates at the
maximum value allowed by the realizability condition for the field strength.
The magnetic field correlation length of a helical magnetic field will also increase more quickly
than for a non-helical field due to the inverse cascade mechanism.
Magnetic fields with maximal helicity are a generic outcome of any extended period of turbulence~\cite{tevzadze}.

Helical magnetic fields can be generated during the electroweak phase
transition or during inflation
\cite{Cornwall:1997ms,Giovannini:1997eg,Field:1998hi,Vachaspati:2001nb,Tashiro:2012mf,
Sigl:2002kt,Subramanian:2004uf,Campanelli:2005ye,semikoz05,DiazGil:2007dy,Campanelli:2008kh,Campanelli:2013mea}.
Such a helical cosmological magnetic field
might be the source of magnetic helicity needed in galactic
dynamo amplification models \cite{Banerjee:2004}. Thus
testing the helicity of any primordial magnetic field is important for
understanding the origin of observed
astrophysical magnetic fields \cite{Widrow}.
Magnetic helicity in strong local magnetic fields like
astrophysical jets can be deduced from the polarization of synchrotron radiation
\cite{deduction1,deduction2}. For cosmological magnetic helicity
the detection issue is more difficult, because the field strengths are
much lower and the observational effects more subtle.

The most direct probe of any cosmological magnetic fields is their effect on the
cosmic microwave background radiation, and particularly its polarization.
The microwave background linear polarization is conventionally decomposed into
E-mode (parity-even) and B-mode (parity-odd) components \cite{kks97,sz97}. A
non-helical magnetic field contributes to all of the parity-even power spectra,
those correlating E with itself and B with itself, in addition to E with the microwave
temperature T and the temperature with itself. These contributions were explicitly
calculated in Ref.~\cite{mack02}, and have been used to constrain the amplitude
of a primordial magnetic field
\cite{Shaw:2009nf,Yamazaki:2011eu,Yamazaki:2010jw,Yamazaki:2010nf,Paoletti:2012bb,Paoletti:2010rx,Kunze:2010ys,Ade:2013zuv}.
However, if a parity-violating helical
magnetic field component is present, then it will contribute to the remaining
parity-odd power spectra, namely EB or TB  \cite{pogosian02,cdk04,kr05,Kunze:2011bp},
which are identically zero for magnetic fields with zero helicity.
Note that Faraday rotation by magnetic
fields \cite{kl96} imprints itself on the power spectrum and frequency spectrum of
microwave background polarization, but
is insensitive to helicity
for a given magnetic field power spectrum
\cite{Ensslin:2003ez,Campanelli:2004pm,Kosowsky:2004zh,kmk08}.

Helical magnetic fields are perhaps the most natural parity-violating source of TB or EB correlations in
the microwave background polarization
\cite{Cabella:2007br,Xia:2007qs,Feng:2006dp,Xia:2012ck,Xia:2009ah,Li:2009rt,Li:2008tma,Gruppuso:2011ci},
but other more speculative parity-violating sources can also induce them.
These include a Chern-Simons coupling of photons to another field
\cite{Lue:1998mq,Feng:2006dp,Cabella:2007br,Xia:2007qs,Xia:2008si,Saito:2007kt},
a homogeneous magnetic field \cite{Scannapieco:1997mt,Scoccola:2004ke,Demianski:2007fz, Kristiansen:2008tx},
Lorentz symmetry breaking
\cite{Carroll:1989vb,Kostelecky:2007zz,Cai:2009uc,Ni:2007ar,Casana:2008ry,Caldwell:2011pu,
MosqueraCuesta:2011tz,Kamionkowski:2010rb,Gluscevic:2010vv,Mewes:2012sm,Gluscevic:2012me,Ni:2009qm,Miller:2009pt,Ni:2009gz},
or non-trivial cosmological topology \cite{Lim:2004js, Carroll:2004ai, Alexander:2006mt, Satoh:2007gn}. If some
non-zero TB or EB correlation is detected, the corresponding angular power spectrum must be measured
sufficiently well to distinguish between these possibilities.

In this paper we obtain upper limits on the helicity of a primordial magnetic field, using
the nine-year WMAP constraints on any cross correlation between microwave background
temperature and B-polarization \cite{komatsu}. Current polarization data is consistent with zero
cosmological TB signal, as expected in the standard cosmological model.
We compute the theoretical estimates of cross correlation
given in Ref. \cite{kr05} and compare with the measured upper limits
\cite{komatsu,Hinshaw:2013,Larson:2010gs}.
Since we obtain only upper limits, we assume that magnetic helicity is the only possible parity-violating
source present, which gives the most conservative helicity upper limits.
For simplicity of calculation, we consider only the vector (vorticity) perturbations
sourced by the magnetic field and neglect the
tensor (gravitational wave) perturbations. This is a good approximation
for angular multipoles $l>50$ \cite{cdk04}, and for this reason we
use measured $C_l^{TB}$ constraints only for $l>50$; the neglected large angular
scales contain little total statistical weight in our constraints.

The outline of the paper is as follows: in Sec.~II we review the main characteristics of a helical magnetic field and
derive the vorticity perturbations. Section III gives the expression for $C_l^{TB}$ due to these vorticity
perturbations, and these are compared with the WMAP 9-year upper limits in Sec.~IV.
Implications and future experimental prospects are discussed in Sec.~V.
We employ natural units with $\hbar = c = 1$ and gaussian units for electromagnetic quantities.

\section{Properties of a Cosmological Magnetic Field}

We assume that a cosmological magnetic field
was generated during or prior to the radiation-dominated epoch,
with the energy density of the field being a first-order
perturbation to the standard Friedmann-Lema\^\i tre-Robertson-Walker
homogeneous cosmological model. We also
assume that primordial plasma is a perfect conductor and thus the spatial
and temporal dependence of the field separates: ${\mathbf
B}({\mathbf x},t)={\mathbf B}({\mathbf x})/a(t)^2$ with $a(t)$ the
cosmological scale factor. The mean helicity density of the magnetic field
is given by
\begin{equation}
{\mathcal H}_B =\frac{1}{V} \int_V d{\bf x} {\bf A({\bf x})}\cdot{\bf B({\bf x})}
= \frac{1}{V}\int_V d{\bf x} {\bf A({\bf x})}\cdot\nabla\times{\bf A({\bf x})} ,
\label{helicity_def}
\end{equation}
with $\bf A$ the vector potential, in the limit that the integral is over an infinite volume.
An integral over a finite but large volume will approximate this helicity density.
In general, magnetic helicity is a gauge-dependent quantity,
because  the vector potential ${\bf A}$ can be redefined by adding a gradient to it.
However, the magnetic helicity is gauge invariant for periodic systems without a net magnetic flux,
as shown in Ref.~\cite{gauge}. We assume that our universe can be well approximated
by a large box with periodic boundary conditions, provided the dimension of the box is
large compared to the Hubble length today. In this case, the magnetic helicity
is a well-defined quantity.

A Gaussian random magnetic field is described by the two-point
correlation function in wavenumber space as
\begin{equation} \label{spectrum}
\langle B^*_m({\mathbf k})B_n({\mathbf k'})\rangle
=(2\pi)^3 \delta^{(3)} ({\mathbf k}-{\mathbf k'})
\left[(\delta_{mn}-\hat{k}_m\hat{k}_n) P_B(k)  + i \epsilon_{mnl}
\hat{k}_l P_H(k)\right] .
\end{equation}
Here $\hat{k}_m=k_m/k$ are the unit wavenumber components,
$\epsilon_{mnl}$ is the antisymmetric tensor, and
 $\delta^{(3)}({\mathbf k}-{\mathbf k'})$ is the Dirac delta
 function. We use the Fourier transform convention
$B_j({\mathbf k}) = \int d^3x \,
 e^{i{\mathbf k}\cdot {\mathbf x}} B_j({\mathbf x})$.
The symmetric power spectrum  $P_B(k)$ is related to the mean magnetic
energy density by
\begin{equation}
{\mathcal E}_B = \frac{1}{(2\pi)^{3}} \int_0^{k_D} dk k^2 P_B(k) ,
\label{energy_density_def}
\end{equation}
while the antisymmetric power spectrum $P_H(k)$ is related
to the magnetic helicity density as
\begin{equation}
{\mathcal H}_B =\frac{1}{(2\pi)^{3}} \int_0^{k_D} dk k \frac{1}{2}P_H(k) ,
\label{helicity_density_def}
\end{equation}
where $k_D$ is a characteristic damping scale for the magnetic field.

The total energy density and helicity of the magnetic field satisfy
the realizability condition
\begin{equation}
{\mathcal H}_{B} \leq 2\xi_M {\mathcal E}_B,
\label{realizability}
\end{equation}
where
\begin{equation}
\xi_M \equiv \frac{2\pi\int_0^{k_D} dk k P_B(k)}{\int_0^{k_D} dk k^2 P_B(k)}
\label{xi_def}
\end{equation}
is the magnetic field correlation length.
The power spectra $P_B(k)$ and $P_H(k)$ are
generically constrained by $P_B(k)\geq |P_H(k)|$.
We assume that these power spectra
are given by simple power laws, $P_B(k)=A_B k^{n_B}$ and
$P_H(k)=A_H k^{n_H}$.
The constraint on their relative amplitudes implies
$n_H> n_B$ \cite{durrer03};
in addition, finiteness of the total magnetic
field energy requires $n_B > -3$ if the power law extends to
arbitrarily small values of $k$.
For physical transparency, instead of describing the
magnetic field amplitude by the proportionality factors $A_B$ and $A_H$,
we will use the effective magnetic field amplitude
$B_{\rm eff} \equiv (8\pi {\mathcal E}_B)^{1/2}$  \cite{ktr11} and the helicity density
${\mathcal H}_B$. Using these quantities is convenient because they do not depend on
the power law indices $n_B$ and $n_H$ and are independent of any smoothing scale.

Often, cosmological magnetic fields are characterized by a smoothed value on
some comoving length scale $\lambda > \lambda_D=2\pi/k_D$.
Convolving with a Gaussian smoothing
kernel, the smoothed magnetic field amplitude $B_\lambda$ is \cite{mack02}
\begin{equation}  \label{energy-spectrum-S}
{B_\lambda}^2 \equiv | \langle{\mathbf
B}({\mathbf x}) \cdot  {\mathbf B}
({\mathbf x}) \rangle|_\lambda
=\frac{2}{(2\pi)^2}A_B\Gamma\left(\frac{n_B+3}{2}\right)\lambda^{-n_B-3},
\qquad\qquad \lambda>\lambda_D.
\end{equation}
We also introduce a smoothed quantity $H_\lambda$
(the so-called {\it helicity measure} or {\it current helicity} \cite{Kunze:2011bp}) related to the magnetic helicity
having the same units as $B_\lambda$ and depending on the
antisymmetric part of the magnetic field spectrum:
\begin{equation}
H_\lambda^2 \equiv \lambda | \langle{\mathbf
B}({\mathbf x}) \cdot [{\mathbf \nabla} \times {\mathbf B}
({\mathbf x})] \rangle|_\lambda
=\frac{2}{(2\pi)^{2}} A_H\Gamma\left(\frac{n_H+4}{2}\right)\lambda^{-n_H-3},
\qquad\qquad\lambda>\lambda_D.
\end{equation}
See Ref.~\cite{Kosowsky:2004zh} for a more detailed discussion.
Then the transformation between the smoothed quantities $B_\lambda$ and $H_\lambda$ and the effective quantities ${ B}_{\rm eff} $ and
${\mathcal H}_B$
 is simply
\begin{equation} \label{eq:beff}
B_{\rm eff} = \frac{B_\lambda (k_D \lambda)^{\frac{n_B + 3}{2}}}
{\sqrt{\Gamma\left(\frac{n_B + 5}{2}\right)}} \kma
\end{equation}
and
\begin{equation}\label{eq:heff}
{\mathcal H}_{B} = \frac{1}{8\pi}
\frac{\lambda H_\lambda^2 (k_D\lambda)^{n_H+2}}{(n_H+2)\Gamma\left(\frac{n_H+4}{2}\right)} \pkt
\end{equation}

We assume that the magnetic field cutoff scale $k_D$ is determined by
the Alfv\'en wave damping scale, $\lambda_D \simeq v_A L_S$ \cite{sub98b,jedamzik98},
where $v_A$ is the Alfv\'en velocity set by
the total magnetic energy density \cite{mack02}. Since $v_A \ll 1$ the Alfv\'en damping scale
will always be much smaller scale than the Silk damping
scale (the thickness of the last scattering surface) for standard
cosmological models.
On the other hand, the CMB fluctuations are determined by the Silk damping scale,
and presence of the magnetic field source at smaller scales will not significantly affect
the resulting spectra.

\section{Microwave Background Fluctuations from a Helical Magnetic Field}

A cosmological magnetic field induces Alfv\'en waves sourced by the Lorentz force in the cosmological plasma
(see \cite{dky98,mack02,jedamzik98,sub98b,Seshadri:2000ky,Subramanian:2003sh}),
which generically produce non-zero vorticity perturbations.
In the case of a stochastic magnetic field the average Lorentz force $\langle{\bf L(x)}\rangle
=-\langle{\bf B} \times [{\bf \nabla} \times {\bf B}]\rangle/(4\pi)$ vanishes,
while the root-mean-square Lorentz force
$\langle{\bf L (x) \cdot L(x)}\rangle^{1/2}$ is non-zero and acts
as a source in the  vector perturbation equation. If the magnetic field
spectrum Eq.~(\ref{spectrum}) has a helical part $P_H(k)$, then the Lorentz force
two-point correlation function will have both symmetric and antisymmetric
pieces. Both contribute to the symmetric piece of the vorticity perturbation
spectrum, but only the antisymmetric piece of the Lorentz force, determined
entirely by $P_H(k)$,  will
contribute to the antisymmetric part of the vorticity perturbation spectrum \cite{kr05}.

In the tight-coupling limit between photons and baryons, the fluid vorticity is sourced
by the transverse and divergence-free piece of the Lorentz force. The fluid vorticity
at last scattering then translates into temperature and polarization fluctuations
in the microwave background radiation \cite{mack02}. The microwave temperature
and E-polarization components are both parity-symmetric, while the B-polarization
component is parity-antisymmetric \cite{hu97}. This implies that the cross-power
spectra $C_l^{TB}$ and $C_l^{EB}$ from stochastic magnetic fields
will be nonzero only if $P_H(k)$ is nonzero \cite{kks97,pogosian02,cdk04,kr05,Kunze:2011bp}.
In other words, the $TB$ and $EB$
power spectra provide a way to measure whether a primordial magnetic
field has a helical component. (A constant magnetic field component also
gives non-zero  $C_l^{TB}$ and $C_l^{EB}$ through Faraday rotation
\cite{Scannapieco:1997mt,Scoccola:2004ke}, but the two distinct contributions can be distinguished by their
different power spectra, and by the frequency dependence of a Faraday rotation
signal.)

Detailed computations of the various CMB angular power spectra induced by
helical and nonhelical magnetic fields have been presented elsewhere \cite{kr05,mack02}. Here
we focus on the TB power spectrum, because current data does not put a significant
constraint on the much smaller EB power spectrum.
For $l>50$ where the TB
power spectrum has significant power, we neglect tensor contributions, which
are smaller. Here we derive an analytic approximation to the TB angular
power spectrum, based on the second-order approximation technique from Ref.~\cite{Zaldarriaga:1995gi};
this approximate solution is simple and accurate enough for deriving upper limits on the helical
magnetic field.

The multipoles of the temperature perturbation from a vector mode in Fourier space are given by
\begin{equation}
\frac{\Theta^{(\pm 1)}_l (k, \eta_0)}{2l+1} \simeq
\sqrt{\frac{l(l+1)}{2}}~\Omega^{(\pm 1)} (k, \eta_{\rm dec}) \frac{j_l (k\eta_0)}{k\eta_0}, \qquad l\geq 2
\label{Theta_def}
\end{equation}
where $\Omega^{\pm 1}(k,\eta)$ are the two helicity components of the
gauge-invariant vorticity perturbations, constructed from the fluid velocity field and the
vector component of the metric perturbations \cite{kr05}.
Here we have made the approximation $\eta_0 -\eta_{\rm dec} \simeq \eta_0$ in Eq.~(\ref{Theta_def}).
For vorticity perturbations sourced by the magnetic field, the $l=1$ moment of temperature fluctuation is well approximated by
the vorticity perturbation, $\Theta^{(\pm 1)}(k, \eta_0) \simeq \Omega^{\pm 1} (k, \eta_{\rm dec})$ \cite{mack02}.
For the B-mode polarization perturbation, we have \cite{mack02}
\begin{equation}
\frac{B^{(\pm 1)}_l (k, \eta_0)}{2l+1}\simeq
\mp \frac{\sqrt{6}}{2} \sqrt{(l-1)(l+2)} \int^{\eta_0}_0 d\eta {\dot\tau}(\eta) e^{-\tau}
P^{(\pm 1)} (k, \eta)\frac{j_l (k\eta_0-k\eta)}{k\eta_0-k\eta}
 \kma
\label{B_def}
\end{equation}
where the polarization source is defined by \cite{hu97}
\be \label{pdef}
P^{(\pm 1)} =\frac{1}{10} \bigl[\Theta^{(\pm 1)}_2 - \sqrt{6} E^{(\pm 1)}_2 \bigr].
\label{polarization_source}
\ee

The temperature and polarization quadrupoles satisfy the evolution equations
\bea
\dot{\Theta}^{(\pm 1)}_2 &=& k \bigl[ \frac{\sqrt{3}}{3}\Theta^{(\pm 1)}_1-\frac{2\sqrt{2}}{7} \Theta^{(\pm 1)}_3\bigr]
+{\dot\tau}\left[ \Theta^{(\pm 1)}_2 - P^{(\pm 1)}\right] \kma \label{Theta_dot}\\
\dot{E}^{(\pm 1)}_2 &=& k \bigl[\mp \frac{1}{3} B^{(\pm 1)}_2 -\frac{2\sqrt{10}}{21} E^{(\pm 1)}_3 \bigr]
+{\dot\tau} \left[E^{(\pm 1)}_2 +\sqrt{6} P^{(\pm 1)} \right]  \label{E_dot}\pkt
\eea
Here the optical depth $\tau(\eta) = \int_\eta^{\eta_0} d \eta^\prime {\dot \tau} (\eta^\prime)$ to photon scattering from conformal time $\eta$ until today satisfies
$d\tau/d\eta \equiv -{\dot\tau}(\eta)=\sigma_T n_e(\eta) a(\eta)$, $\sigma_T$ is the
Thomson scattering cross section, and $n_e(\eta)$ the comoving number density of free electrons.
The vector mode of the CMB temperature-$B$ polarization angular power spectrum is given by \cite{hu97}
\begin{equation}
C^{TB~(V)}_l = \frac{2}{\pi} \int dk k^2 \left[ \frac{\Theta^{(-1)*}_l (k, \eta_0)}{2l+1} \frac{B^{(-1)}_l (k, \eta_0)}{2l+1}
+ \frac{\Theta^{(+1)*}_l (k, \eta_0)}{2l+1} \frac{B^{(+1)}_l (k, \eta_0)}{2l+1} \right] \pkt
\end{equation}
In the rest of this section, we approximate this power spectrum in a given cosmological model, for comparison with
limits on this power spectrum from temperature and polarization sky maps.

The visibility function ${\dot\tau} e^{-\tau}$
is sharply peaked at the time of decoupling, so to determine the $B$-polarization signal, Eq. (\ref{B_def}), we need to know the polarization source $P^{(\pm 1)}$ at the time of decoupling.
Differentiating Eq.~(\ref{polarization_source}) with respect to conformal time and substituting
Eqs.~(\ref{Theta_dot}) and (\ref{E_dot}), at leading order we get
\be \label{peq}
\dot{P}^{(\pm 1)} -\frac{3}{10} {\dot\tau} P^{(\pm 1)} \simeq \frac{k \sqrt{3}}{30} \Theta^{(\pm 1)}_1 \kma
\ee
where we have dropped terms containing $\Theta^{(\pm 1)}_3, E^{(\pm 1)}_3$ and $B^{(\pm 1)}_2$
(see also Ref. \cite{Zaldarriaga:1995gi}).

In our previous work \cite{kr05,mack02}, we assumed the first term of Eq. (\ref{peq}) is small to obtain the
approximate solution
$P^{(\pm 1)} = \sqrt{3}k\Theta_1^{(\pm 1)}/9{\dot\tau}$. While usually valid, this approximation
fails during recombination because $k/\dot \tau $ varies rapidly:
inserting
$P^{(\pm 1)} = \sqrt{3}k\Theta_1^{(\pm 1)}/9{\dot\tau}$ into the integral of Eq. (\ref{B_def}), the integrand becomes proportional to $e^{-\tau}$,
and is not anymore peaked at the time of decoupling.
Instead, we employ
a more precise second-order approximate solution to the source equation, following the technique in Refs.~\cite{Zaldarriaga:1995gi,Cai:2012ci}.
Details are given in the Appendix; the solution for the temperature-B polarization power spectrum is,
\begin{eqnarray}
C_l^{TB} &\simeq& -\frac{3\pi}{14} \ln\left(\frac{10}{3}\right)
\sqrt{\frac{(l+2)!}{(l-2)!}} \frac{(n_B+3)(n_H+2)}{n_B+n_H+2} \frac{\eta_{\rm dec}^2}{\eta_0^2}{k_D\Delta\eta_{\rm dec}}
\frac{1}{(1+R_{\rm dec})^2}\nonumber\\
&&\times\frac{{\cal E}_B{\cal H}_B k_D}{\rho_{\gamma\,0}^2}
 \int_0^{x_S}dx\,x^4 D_E(x)
\left[1+\frac{n_H-1}{n_B+3}
x^{n_B+n_H+2} \right] j_{l}^2(xk_D\eta_0)
\label{ClTB}
\end{eqnarray}
with the change of variables $x=k/k_D$ in the integral. We have defined a function which models the effect of
Silk damping for polarization \cite{Cai:2012ci},
\begin{equation}
D_E(x) \equiv 0.2 \left( e^{-c_E (a_1 x k_D\eta_0)^{b_E}} + e^{-c_E (a_2 x k_D\eta_0)^{b_E}}\right)
\label{D_E_def}
\end{equation}
with the fitting constants $c_E=0.27$, $b_E=2.0$, $a_1 = 0.0011$, and $a_2=0.0019$.
The amplitude of the approximate solution Eq.~(\ref{ClTB})
differs from that in Ref.~\cite{mack02} by roughly a factor of two.

\section{Constraints from WMAP}

We obtain constraints on primordial magnetic helicity by comparing the
temperature-$B$-polarization cross correlation function in Eq.~(\ref{ClTB})
with WMAP nine-year data. We assume a standard $\Lambda$CDM model.
We take the Silk damping scale to be the thickness of the last scattering surface,
$L_S \simeq \Delta \eta_{\rm dec}$, which is determined by  the function $D_E(x)$,
so $k_S = 0.3 $~Mpc$^{-1}$.
The WMAP $C_l^{TB}$ measurement is consistent with
a null signal, as expected in the standard cosmological model.
We follow a Feldman-Cousins prescription~\cite{Feldman:1997qc}
to set $68\%$ and $95\%$ confidence level (CL) upper limits
on the primordial magnetic field \cite{komatsu}.
We only consider multipoles with $\ell > 50$ to simplify the analysis;
in this range the measured values of $C_\ell^{TB}$
are uncorrelated between different $\ell$ values.
This restriction does not significantly impact sensitivity to
the magnetic field, since most signal is for larger values of multipole number.

A comparison between our model for $C_\ell^{TB}$ and the nine-year WMAP data is given in Fig.~\ref{fig:model} for
two magnetic field helicity models: one with power law $n_H = -1.9$ and amplitude ${\mathcal H}_{B} = 10^5$ nG$^2$ Mpc,
and one with power law $n_H = -0.6$ and amplitude ${\mathcal H}_{B} = 10^8$ nG$^2$ Mpc. For both cases, we set the value of the effective
magnetic field $B_{\rm eff}$ to 1 nG and the spectral index to its inflationary value of $n_B = -2.99$, which is
somewhat below current cosmological limits \cite{current_limits}.
These models both produce a helical magnetic field
which is just at the level which can be ruled out
from the WMAP 9-year microwave background polarization power spectra. The helicity amplitude ${\mathcal H}_{B}$ varies strongly with spectral index, because
for larger values of $n_H$ the helicity is more concentrated on small scales, close to the damping scale, which contribute
little to the microwave background signal.

\begin{figure}[htbp]
\begin{center}
\includegraphics[width=0.8\textwidth]{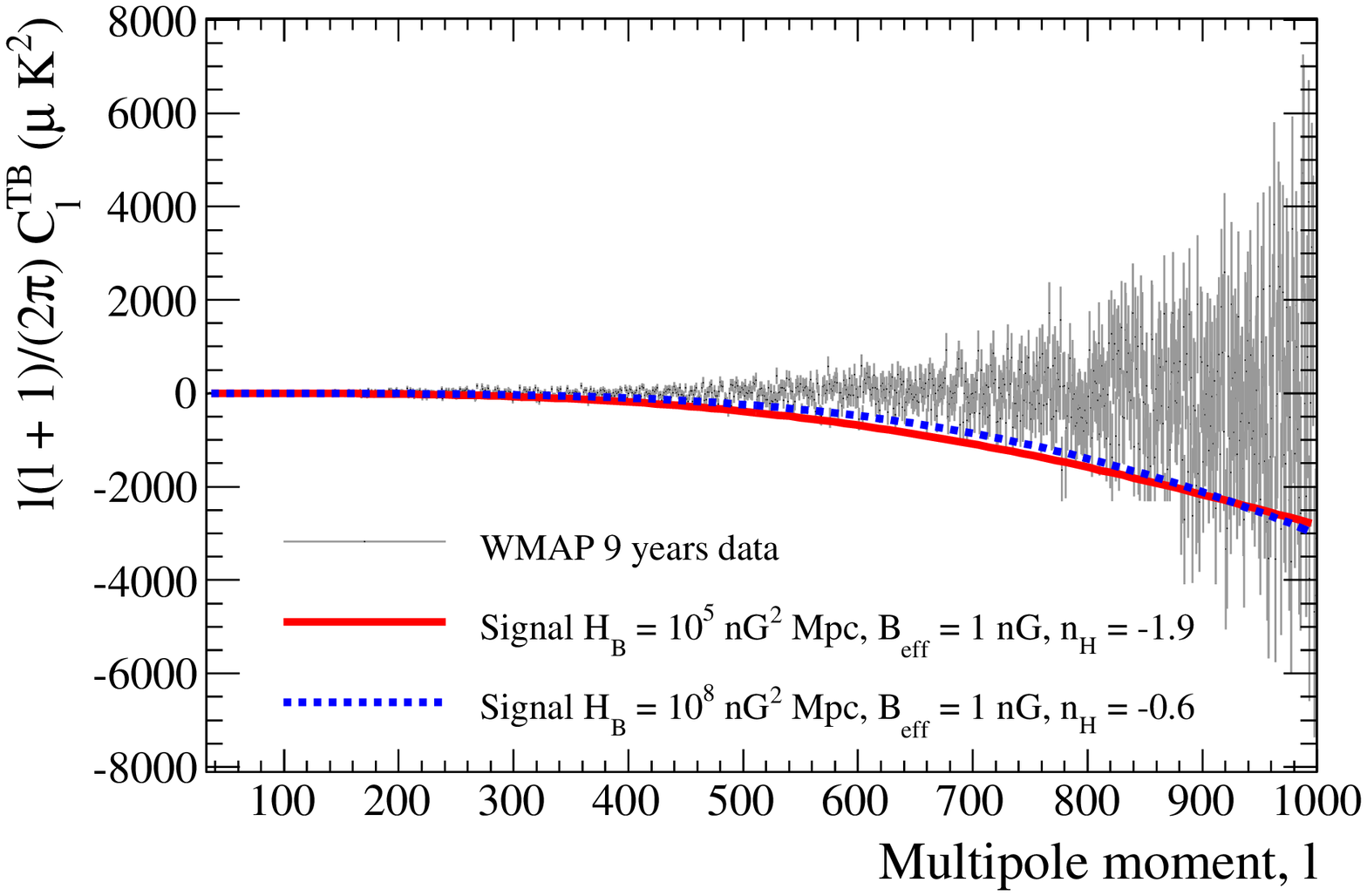}
\labelcaption{fig:model}
{A comparison between the temperature-$B$-polarization cross correlation model
for $B_{\rm eff}$ = 1 nG, $n_B = -2.99$; the solid red line is for a helicity amplitude of ${\mathcal H}_{B} = 10^5$ nG$^2$ Mpc and
helicity spectral index $n_H = -1.9$ while the dotted blue line is for
${\mathcal H}_{\rm B} = 10^8$ nG$^2$ Mpc and $n_H$ = -0.6. Also shown are the nine-year WMAP
data (solid gray dots with bars indicating uncertainties).}
\end{center}
\end{figure}

The upper limits on the ${\mathcal H}_{B}$ as functions of $n_H$ are given in Fig.~\ref{fig:heff} for three
scenarios: $n_B = -2.99$, $n_B = -2.0$, and $n_B = n_H - 1$.
\begin{figure}[htbp]
\begin{center}
\includegraphics[width=0.8\textwidth]{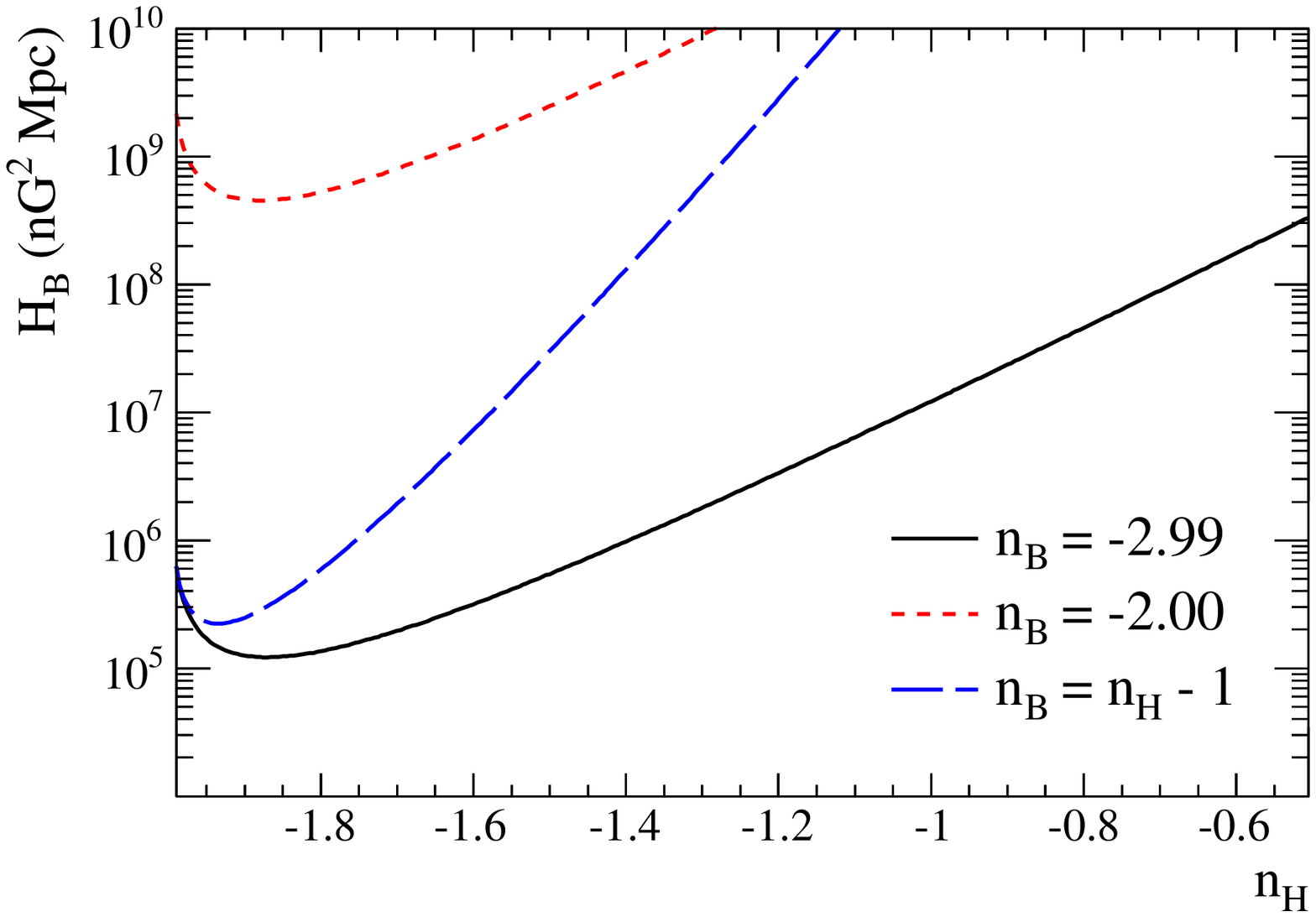}
\labelcaption{fig:heff}
{The 95\% upper limits on ${\mathcal H}_{B}$ as a function of $n_H$ for
$n_B = -2.99$ (black solid), $n_B = -2.0$ (red short dash), and $n_B = n_H - 1$ (blue long dash). For all three cases,
$B_{\rm eff} =1$ nG.}
\end{center}
\end{figure}
We also present the limits in terms of $H_\lambda$ for the same three scenarios in Fig.~\ref{fig:hlambda}, using
a smoothing scale of $\lambda = 1$ Mpc which is commonly used in the magnetic field literature.
\begin{figure}[htbp]
\begin{center}
\includegraphics[width=0.8\textwidth]{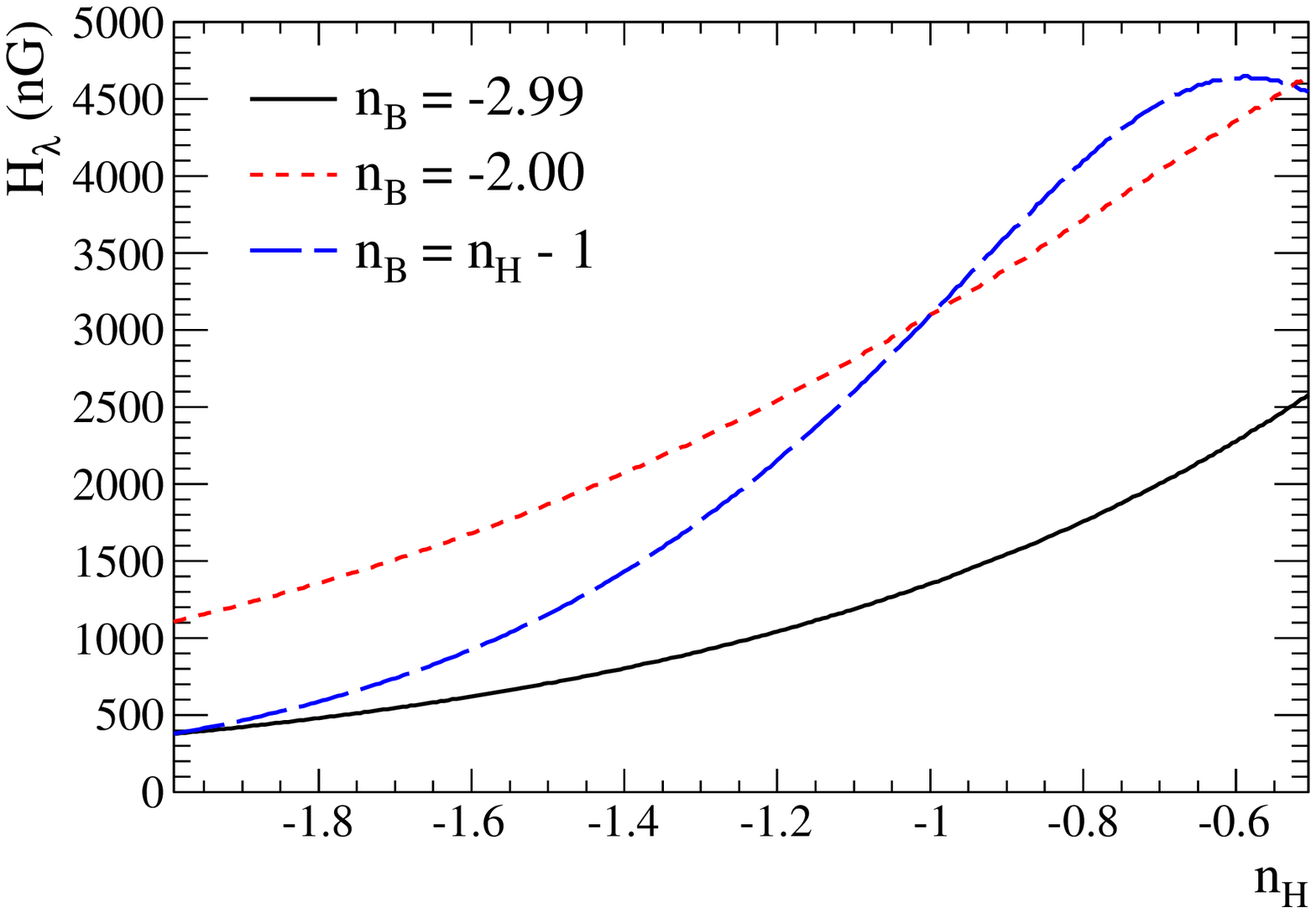}
\labelcaption{fig:hlambda}
{The 95\% upper limits on ${\mathcal H}_\lambda$ for a smoothing scale $\lambda=1$ Mpc, as a function of $n_H$ for
$n_B = -2.99$ (black solid), $n_B = -2.0$ (red short dash), and $n_B = n_H - 1$ (blue long dash). For all three cases,  $B_{\rm eff}=1$ nG.}
\end{center}
\end{figure}

The results are relatively insensitive to the systematic uncertainty in the cross-correlation signal due to modeling of the cutoff scale;
a plausible range of cutoff scales gives a signal difference which is smaller than the measurement uncertainties in the WMAP data.
Systematic uncertainties with a size up to 20\% of the predicted values of $C_l^{TB}$ have only small effects on the magnetic
field limits obtained here.

\section{Conclusions}

The results presented here are the first direct constraint on a helical primordial magnetic field by its
contribution to the parity-odd temperature-B polarization cross-power spectrum $C_l^{TB}$ of the microwave background.
No experiment to date has detected a non-zero value for this power spectrum; we use the WMAP 9-year measurement
which is consistent with zero to place upper limits on the combined mean field strength and helicity of a primordial
magnetic field.
The primordial magnetic field amplitude constraint of around $B_{\rm eff} = 3$ nG from
the microwave background temperature and E-polarization power spectra \cite{Yamazaki:2013hda,Shaw:2010ea,Ade:2013zuv,Kahniashvili:2012dy}
gives an upper limits on magnetic helicity ${\mathcal H}_B$ less
than around 10 nG$^2$ Gpc for a nearly scale-invariant power spectrum with $n_B=-2.99$.
The helicity limits become weaker for larger values of $n_B$. Recent work has argued for more stringent upper limits of
$B_{\rm eff} < 1$ nG from constraints on the trispectrum induced by magnetic fields, rather than the power spectrum \cite{trivedi14}.
If magnetic fields from inflation are produced with a magnetic curvature mode as advocated by Ref.~\cite{bonvin13}, then the trispectrum
constraint is even stronger, pushing the magnetic field amplitude down to $B_{\rm eff} < 0.05$ nG.

The mean helicity amplitude over a given volume is constrained by the realizability condition Eq.~(\ref{realizability}). The smaller
the value of the magnetic field $B_{\rm eff}$, the lower the helicity ${\mathcal H}_B$ which can be supported by the field.
Any cosmological field will have physical effects measured over an effective volume which is at most the Hubble volume,
so the effective comoving correlation length of this field is limited by the Hubble length $H_0^{-1}$. For a given microwave
background constraint on ${\mathcal H}_B$ and assuming a magnetic field strength equal to some current upper limit,
the maximal magnetic
helicity which saturates the realizability condition must have a correlation length $\xi_M = 4\pi {\mathcal H}_B/B_{\rm eff}^2$.
If this correlation length is larger than the Hubble length, then a magnetic field of the given amplitude cannot support
helicity as large as the measured limit. For a magnetic field with $B_{\rm eff} = 3$ nG and the corresponding helicity
equal to the limiting value ${\mathcal H}_B=10$ nG$^2$ Gpc , the correlation length for
maximal helicity is around 10 Gpc: current measurements provide a helicity constraint which is just at the level
of the maximum possible helicity for the magnetic field strength. If the field strength is significantly lower, then
the helicity limits derived in this paper are substantially above the maximum helicity allowed by Eq.~(\ref{realizability}).

Upcoming polarization data from the Planck satellite, as
well as high-resolution ground-based experiments like ACTPol \cite{actpol} and  SPTPol \cite{sptpol},
 will strengthen limits on both the magnetic field amplitude and helicity, for
 two reasons:
first, the signal increases for larger $l$ values beyond those probed by WMAP, and second,
upcoming experiments will produce polarized maps over large portions of the sky with much greater
sensitivity than WMAP. Interest in B-mode polarization has exploded due to the recent results from the BICEP2
collaboration \cite{bicep2}. Experiments searching for B-polarization from primordial tensor modes
(at large angular scales) and gravitational lensing (at small angular scales)
will drive continual increases in sensitivity over the coming decade.
Planck's maps have a sensitivity (around 85 $\mu$K-arcmin for the SMICA map)
which is a factor of 4 lower than WMAP (around 360 $\mu$K-arcmin), corresponding to
errors in $C_l^{TB}$ smaller by a factor of 16. The recent PRISM satellite proposal \cite{prism}
envisions full-sky polarization maps with sensitivity of 3 $\mu$K-arcmin, which would
give $C_l^{TB}$ errors smaller than the WMAP errors used here by a factor of $10^4$.

Limits on the magnetic field amplitude $B_{\rm eff}$ from the microwave background power
spectra will not improve
substantially, because they are limited by cosmic variance in the power spectra from other
non-magnetic sources of fluctuations. In contrast, sensitivity improvements in polarization will
continue to  improve helicity limits from $C_l^{TB}$ because this signal is not limited
by cosmic variance: it is zero for standard-cosmology primary perturbations which do not
violate parity. (At least this is the case until extreme sensitivities are reached where the cosmic variance in
$C_l^{TB}$ from the residual gravitational lensing contribution to de-lensed maps dominates
over the map noise). So future measurements may provide constraints on magnetic
field helicity which are much below the maximal helicity allowed by Eq.~(\ref{realizability}) and
the magnetic field amplitude limits.

The $TB$ power spectrum of cosmic microwave background polarization, and its
lower-amplitude counterpart $EB$, provide a valuable opportunity to probe
unconventional physics which violates cosmological parity. Of contributors
to these power spectra, gravitational lensing and helical magnetic fields are the
two sources which rely only on standard, demonstrated physical effects.
The microwave background
lensing spectrum can be calculated to high accuracy within the standard model
of cosmological structure formation, so any departures from this signal would
be a good bet for revealing the existence of significant helical magnetic
fields in the universe. In turn, the detection of helicity would
give valuable information about the still-mysterious origin of magnetic
field in the cosmos.

\acknowledgments
It is our pleasure to thank A. Brandenburg, L. Campanelli, K. Kunze, H. Tashiro, A.~Tevzadze,  and T.~Vachaspati for useful discussions.
T.K.\ and G.L.\ acknowledge partial support from the Swiss NSF SCOPES grant IZ7370-152581.
T.K.\ and A.K.\ were supported in part through NASA Astrophysics Theory  program grant NNXl0AC85G and NSF Astrophysics and Astronomy Grant Program grants AST-1109180 and AST-1108790. T.K. acknowledges partial support from Berkman Foundation. A.K. is partly supported by NSF Astrophysics and Astronomy grant AST-1312380.
This work made use of the NASA Astrophysical Data System for bibliographic information.

\appendix
\section{Derivation of $C^{TB~(V)}_l$ }

The solution of Eq.~(\ref{peq}) can be written in the form
\bea \label{p_solution}
P^{(\pm 1)}=\frac{\sqrt{3}}{30} k \int^\eta_0 d\eta' e^{-\frac{3}{10} \tau(\eta, \eta')} \Theta^{(\pm 1)}_1 (\eta')
= \frac{\sqrt{3}}{30} k \int^\eta_0 d\eta' e^{+\frac{3}{10} \tau(\eta)}
e^{-\frac{3}{10} \tau(\eta')} \Theta^{(\pm 1)}_1 (\eta') \kma
\eea
where $\tau(\eta, \eta')=\int^\eta_{\eta'} d{\eta}'' {\dot\tau}(\eta'')=\tau(\eta')-\tau(\eta)$ and
the visibility function $V(\eta)=\frac{d}{d\eta}e^{-\tau(\eta)} = {\dot\tau} e^{-\tau}$ can be approximated by the
asymmetric Gaussian function \cite{gorbunov,Cai:2012ci}
\be
V(\eta)=V(\eta_{\rm dec}) {\rm exp} \bigl[-\frac{(\eta-\eta_{\rm dec})^2}{2\Delta\eta_{\rm dec}^2} \bigr] \kma
\ee
where $\Delta\eta_{\rm dec}= \Delta\eta_{{\rm dec}_1} \Theta(\eta_{\rm dec}-\eta)+ \Delta\eta_{{\rm dec}_2}
\Theta(\eta-\eta_{\rm dec})$,
$\Delta\eta_{{\rm dec}_1}=0.0011 \eta_0$ and $\Delta\eta_{{\rm dec}_2}=0.0019\eta_0$ and $\Theta(\eta)$ is the usual step
function. The prefactor $V(\eta_{\rm dec})=\frac{1}{\sqrt{2\pi} \Delta\eta_{\rm dec}}$ is calculated from the
normalization condition $\int^{\eta_0}_0 V(\eta)d\eta=1$.
Substituting the solution Eq.~(\ref{p_solution}) into Eq.~(\ref{B_def}) we obtain
\be \label{Bl_2}
\frac{B^{\pm 1}_l (k, \eta_0)}{2l+1} =
\mp \frac{k}{10 \sqrt{2}} \sqrt{(l-1)(l+2)} \int^{\eta_0}_0 d\eta V(\eta)
\frac{j_l (k(\eta_0 - \eta))}{k(\eta_0 - \eta)}
  \int^\eta_0 d\eta' e^{+\frac{3}{10} \tau(\eta)}
e^{-\frac{3}{10} \tau(\eta')} \Theta^{(\pm 1)}_1 (\eta') \pkt
\ee
Since the visibility function $V(\eta)$ is sharply peaked around $\eta=\eta_{\rm dec}$ and
$e^{-\frac{3}{10} \tau(\eta')}$ behaves like a step function, the $\Theta^{(\pm 1)}_1 (\eta')$ factor
can approximately be pulled out from the $\eta'$ integration and we get
\be \label{Bl_3}
\frac{B^{\pm 1}_l (k, \eta_0)}{2l+1} =
\mp \frac{k}{10 \sqrt{2}} \sqrt{(l-1)(l+2)} \int^{\eta_0}_0 d\eta V(\eta)
\frac{j_l (k(\eta_0 - \eta))}{k(\eta_0 - \eta)} \Theta^{(\pm 1)}_1 (\eta)
\int^\eta_0 d\eta' e^{+\frac{3}{10} \tau(\eta)} e^{-\frac{3}{10} \tau(\eta')}  \pkt
\ee

Noticing that $V(\eta)\propto \exp(-\gamma (\eta-\eta_{\rm dec})^2)$ and $j_l (k(\eta_0-\eta))$ contains
a mixture of oscillating modes $e^{ip\eta}$ and $e^{-ip\eta}$ with $p\propto k$, the formula
$\int_{-\infty}^{\infty} e^{-\gamma \eta^2}e^{ip\eta}d\eta =
e^{-p^2/4\gamma} \int_{-\infty}^{\infty} e^{-\gamma \eta^2} d\eta$
gives the approximation \cite{Cai:2012ci}
\be
\int^{\eta_0}_0 d\eta V(\eta) \frac{j_l (k(\eta_0 - \eta))}{k(\eta_0 - \eta)} \Theta^{(\pm 1)}_1 (\eta) \approx
\frac{j_l (k(\eta_0-\eta_{\rm dec})}{k \eta_0 - k\eta_{\rm dec}} \Theta^{(\pm 1)}_1 (\eta_{\rm dec}) D_E (k)
\int^{\eta_0}_0 d\eta V(\eta) \kma
\ee
where $D_E(k)$ is the Silk damping factor for polarization \cite{Cai:2012ci}, Eq.~(\ref{D_E_def}).

Introducing a new variable $x\equiv \tau(\eta')/\tau(\eta)$, approximating
$d\eta' = -\Delta \eta_{\rm dec} dx/x $
and noticing that
\be
\int^{\eta_0}_0 d\eta ~ V(\eta) \int^{\infty}_1 \frac{dx}{x} e^{-\frac{3}{10}x \tau(\eta)} e^{\frac{3}{10} \tau(\eta)}
= \Delta \eta_{\rm dec} \cdot \int^{\infty}_0 d\tau e^{-\frac{7}{10}\tau(\eta)} \int^{\infty}_1 \frac{dx}{x}
e^{-\frac{3}{10} x \tau(\eta)}
=\Delta \eta_{\rm dec} \frac{10}{7} {\rm ln} \frac{10}{3} \kma
\ee
we get
\bea
\frac{B^{\pm 1}(k, \eta_0)}{2l+1} &=&\mp \frac{\sqrt{2}}{14} {\rm ln}(\frac{10}{3}) \sqrt{(l-1)(l+2)} D_E (k)
\frac{j_l (k\eta_0)}{k\eta_0} k \Delta\eta_{\rm dec} \Theta^{(\pm 1)}_1 (k, \eta_{\rm dec})
\\ \label{B_fin}
&=& \mp\frac{\sqrt{2}}{14} {\rm ln}(\frac{10}{3}) \sqrt{(l-1)(l+2)}  D_E (k) k \Delta\eta_{\rm dec}
\frac{j_l (k\eta_0)}{k\eta_0} \Omega^{(\pm 1)} (k, \eta_{\rm dec}) \pkt
\eea
Making use of Eq.~(\ref{B_fin}) and Eq.~(\ref{Theta_def}),
we finally obtain for the temperature-$B$-polarization cross correlation function
\be \label{A9}
C^{TB~(V)}_l = - \frac{2}{7 \pi} {\rm ln}(\frac{10}{3}) \sqrt{\frac{(l+2)!}{(l-2)!}} \int dk k^2 D_E(k)
\omega(k, \eta_{\rm dec}) \frac{j_l^2(k\eta_0)}{(k\eta_0)^2} k \Delta \eta_{\rm dec} \kma
\ee
where $\omega(k)$ is the helical part of the power spectrum which can be expressed as \cite{kr05}
\begin{equation} \label{A10}
\omega(k, \eta) = \left[\frac{k\eta}{(\rho_{\gamma, 0} + p_{\gamma, 0})(1+R_{\rm dec})}\right]^2 g(k).
\label{omegak}
\end{equation}
Here $p_{\gamma 0}$ and $\rho_{\gamma 0}$ are the radiation pressure and energy density today,
$R_{\rm dec}$ is the baryon-photon energy density at decoupling, and $g(k)$ can be expressed in terms of
the spectral indices $n_B$ and $n_H$,
values of $B_\lambda$, $H_\lambda$, and the smoothing scale $\lambda$ as
\begin{equation}\label{gk}
g(k) = {\mathcal G} \lambda k (\lambda k_D)^{n_B+n_H+2}
\left[1+\frac{n_H-1}{n_B+3}\left(\frac{k}{k_D}\right)^{n_B+n_H+2}\right]
\end{equation}
with
\begin{equation}\label{G}
{\mathcal G} = \frac{\lambda^3 B_\lambda^2 H_\lambda^2}{24(n_B+n_H+2)\Gamma\left(\frac{n_B+3}{2}\right)
\Gamma\left(\frac{n_H+4}{2}\right)}
\end{equation}
Then using Eqs.~(\ref{A10}), (\ref{gk}), and (\ref{G})
in Eq.~(\ref{A9}) we arrive at Eq.~(\ref{ClTB}).



\begin{thebibliography}{}

\bibitem{Widrow}
L.~M.~Widrow,
``Origin of galactic and extragalactic magnetic fields,''
Rev.\ Mod.\ Phys.\  {\bf 74}, 775 (2002)
[arXiv:astro-ph/0207240].

\bibitem{Kulsrud:2007an}
R.~M.~Kulsrud and E.~G.~Zweibel,
``The Origin of Astrophysical Magnetic Fields,''
Rept.\ Prog.\ Phys.\  {\bf 71}, 0046091 (2008)
[arXiv:0707.2783 [astro-ph]].

\bibitem{Kandus:2010nw}
A.~Kandus, K.~E.~Kunze and C.~G.~Tsagas,
``Primordial magnetogenesis,''
Phys.\ Rept.\  {\bf 505}, 1 (2011).
[arXiv:1007.3891 [astro-ph.CO]].

\bibitem{Durrer:2013pga}
R.~Durrer and A.~Neronov,
  ``Cosmological Magnetic Fields: Their Generation, Evolution and Observation,''
Astron. Astrophys. Rev {\bf 21}, 62 (2013).
  [arXiv:1303.7121 [astro-ph.CO]].

\bibitem{Brandenburg:2004jv}
A.~Brandenburg and K.~Subramanian,
``Astrophysical magnetic fields and nonlinear dynamo theory,''
Phys.\ Rept.\  {\bf 417}, 1 (2005)  [arXiv:astro-ph/0405052].

\bibitem{tevzadze}
A.~G.~Tevzadze, L.~Kisslinger, A.~Brandenburg and T.~Kahniashvili,
  ``Magnetic Fields from QCD Phase Transitions,''
  Astrophys.\ J.\  {\bf 759}, 54 (2012)
  [arXiv:1207.0751 [astro-ph.CO]].

\bibitem{Cornwall:1997ms}
J.~M.~Cornwall,
``Speculations on primordial magnetic helicity,''
Phys.\ Rev.\  D {\bf 56}, 6146 (1997)
[arXiv:hep-th/9704022].

\bibitem{Giovannini:1997eg}
M.~Giovannini and M.~E.~Shaposhnikov,
``Primordial hypermagnetic fields and triangle anomaly,''
Phys.\ Rev.\  D {\bf 57}, 2186 (1998)
[arXiv:hep-ph/9710234].

\bibitem{Field:1998hi}
G.~B.~Field and S.~M.~Carroll,
``Cosmological magnetic fields from primordial helicity,''
Phys.\ Rev.\  D {\bf 62}, 103008 (2000)
[arXiv:astro-ph/9811206].

\bibitem{Vachaspati:2001nb}
T.~Vachaspati,
``Estimate of the primordial magnetic field helicity,''
Phys.\ Rev.\ Lett.\  {\bf 87}, 251302 (2001)
[arXiv:astro-ph/0101261].

\bibitem{Tashiro:2012mf}
  H.~Tashiro, T.~Vachaspati and A.~Vilenkin,
  ``Chiral Effects and Cosmic Magnetic Fields,''
  Phys.\ Rev.\ D {\bf 86}, 105033 (2012)
  [arXiv:1206.5549 [astro-ph.CO]].

\bibitem{Sigl:2002kt}
G.~Sigl,
``Cosmological Magnetic Fields from Primordial Helical Seeds,''
Phys.\ Rev.\  D {\bf 66}, 123002 (2002)
[arXiv:astro-ph/0202424].

\bibitem{Subramanian:2004uf}
K.~Subramanian and A.~Brandenburg,
``Nonlinear current helicity fluxes in turbulent dynamos and alpha quenching,''
Phys.\ Rev.\ Lett.\  {\bf 93}, 205001 (2004)
[arXiv:astro-ph/0408020].

\bibitem{Campanelli:2005ye}
L.~Campanelli and M.~Giannotti,
``Magnetic helicity generation from the cosmic axion field,''
Phys.\ Rev.\  D {\bf 72}, 123001 (2005)
[arXiv:astro-ph/0508653].

\bibitem{semikoz05}
V.~B.~Semikoz and D.~D.~Sokoloff,
``Magnetic helicity and cosmological magnetic field,''
arXiv:astro-ph/0411496;
``Large-scale cosmological magnetic fields and magnetic helicity,''
Int.\ J.\ Mod.\ Phys.\  D {\bf 14}, 1839 (2005).

\bibitem{DiazGil:2007dy}
A.~Diaz-Gil, J.~Garcia-Bellido, M.~Garcia Perez and A.~Gonzalez-Arroyo,
``Magnetic field production during preheating at the electroweak scale,''
Phys.\ Rev.\ Lett.\  {\bf 100}, 241301 (2008)
[arXiv:0712.4263 [hep-ph]].

\bibitem{Campanelli:2008kh}
L.~Campanelli,
``Helical Magnetic Fields from Inflation,''
Int.\ J.\ Mod.\ Phys.\ D {\bf 18}, 1395 (2009)  [arXiv:0805.0575 [astro-ph]].

\bibitem{Campanelli:2013mea}
 L.~Campanelli,
 ``Origin of Cosmic Magnetic Fields,''
 Phys.\ Rev.\ Lett.\  {\bf 111}, no. 6, 061301 (2013)
 [arXiv:1304.6534 [astro-ph.CO]].

\bibitem{Banerjee:2004}
R.~Banerjee and K.~Jedamzik,
``The evolution of cosmic magnetic fields: From the very early universe, to
recombination, to the present,''
Phys.\ Rev.\  D {\bf 70}, 123003 (2004)
[arXiv:astro-ph/0410032].


\bibitem{deduction1}
T.~A.~Ensslin,
``Does circular polarisation reveal the rotation of quasar engines?,''
Astron.\ Astrophys.\  {\bf 401}, 499 (2003)
[arXiv:astro-ph/0212387].

\bibitem{deduction2}
J.P.~Vall{\'e}e,
New Astron.\ Rev.\ {\bf 48}, 763 (2004).

\bibitem{kks97}
M.~Kamionkowski, A.~Kosowsky, and A.~Stebbins,
``Statistics of Cosmic Microwave Background Polarization,''
Phys.\ Rev.\ D {\bf 55}, 7368 (1997) [arXiv:astro-ph/9611125].

\bibitem{sz97}
 M.~Zaldarriaga and U.~Seljak,
 ``An all sky analysis of polarization in the microwave background,''
 Phys.\ Rev.\ D {\bf 55}, 1830 (1997)
 [arXiv:astro-ph/9609170].

\bibitem{mack02}
A.~Mack, T.~Kahniashvili and A.~Kosowsky,
``Vector and Tensor Microwave Background Signatures of a Primordial
Stochastic Magnetic Field,''
Phys.\ Rev.\  D {\bf 65}, 123004 (2002) [arXiv:astro-ph/0105504].

\bibitem{Shaw:2009nf}
  J.~R.~Shaw and A.~Lewis,
  ``Massive Neutrinos and Magnetic Fields in the Early Universe,''
  Phys.\ Rev.\ D {\bf 81}, 043517 (2010)
  [arXiv:0911.2714 [astro-ph.CO]].

\bibitem{Yamazaki:2011eu}
  D.~G.~Yamazaki, K.~Ichiki, T.~Kajino and G.~J.~Mathew,
  ``Primordial Magnetic Field Effects on the CMB and Large Scale Structure,''
  Adv.\ Astron.\  {\bf 2010}, 586590 (2010)
  [arXiv:1112.4922 [astro-ph.CO]].

\bibitem{Yamazaki:2010jw}
  D.~G.~Yamazaki, K.~Ichiki, T.~Kajino and G.~.J.~Mathews,
  ``Constraints on the neutrino mass and the primordial magnetic field from the matter density fluctuation parameter $\sigma_8$,''
  Phys.\ Rev.\ D {\bf 81}, 103519 (2010)
  [arXiv:1005.1638 [astro-ph.CO]].

\bibitem{Yamazaki:2010nf}
  D.~G.~Yamazaki, K.~Ichiki, T.~Kajino and G.~J.~Mathews,
  ``New Constraints on the Primordial Magnetic Field,''
  Phys.\ Rev.\ D {\bf 81}, 023008 (2010)
  [arXiv:1001.2012 [astro-ph.CO]].


\bibitem{Paoletti:2012bb}
  D.~Paoletti and F.~Finelli,
  ``Constraints on a Stochastic Background of Primordial Magnetic Fields with WMAP and South Pole Telescope data,''
  Phys.\ Lett.\ B {\bf 726}, 45 (2013)
  [arXiv:1208.2625 [astro-ph.CO]].

\bibitem{Paoletti:2010rx}
  D.~Paoletti and F.~Finelli,
  ``CMB Constraints on a Stochastic Background of Primordial Magnetic Fields,''
  Phys.\ Rev.\ D {\bf 83}, 123533 (2011)
  [arXiv:1005.0148 [astro-ph.CO]].

\bibitem{Kunze:2010ys}
K.~E.~Kunze,
``CMB anisotropies in the presence of a stochastic magnetic field,''
Phys.\ Rev.\ D {\bf 83}, 023006 (2011)  [arXiv:1007.3163 [astro-ph.CO]].

\bibitem{Ade:2013zuv}
  P.~A.~R.~Ade {\it et al.}  [Planck Collaboration],
  ``Planck 2013 results. XVI. Cosmological parameters,'' preprint (2013)
  [arXiv:1303.5076 [astro-ph.CO]].

\bibitem{pogosian02}
L.~Pogosian, T.~Vachaspati and S.~Winitzki,
``Signatures of kinetic and magnetic helicity in the CMBR,''
Phys.\ Rev.\  D {\bf 65}, 083502 (2002)
[arXiv:astro-ph/0112536].

\bibitem{cdk04}
C.~Caprini, R.~Durrer and T.~Kahniashvili,
``The cosmic microwave background and helical magnetic fields: The tensor mode,''
Phys.\ Rev.\  D {\bf 69}, 063006 (2004)
[arXiv:astro-ph/0304556].

\bibitem{kr05}
T.~Kahniashvili and B.~Ratra,
``Effects of cosmological magnetic helicity on the cosmic microwave background,''
Phys.\ Rev.\  D {\bf 71}, 103006 (2005)
[arXiv:astro-ph/0503709].

\bibitem{Kunze:2011bp}
K.~E.~Kunze,
``Effects of helical magnetic fields on the cosmic microwave background,''
Phys.\ Rev.\ D {\bf 85}, 083004 (2012)  [arXiv:1112.4797 [astro-ph.CO]].

\bibitem{kl96}
 A.~Kosowsky and A.~Loeb,
 ``Faraday rotation of microwave background polarization by a primordial magnetic field,''
 Astrophys.\ J.\  {\bf 469}, 1 (1996);
 [arXiv:astro-ph/9601055].

\bibitem{Ensslin:2003ez}
T.~A.~Ensslin and C.~Vogt,
``The Magnetic Power Spectrum in Faraday Rotation Screens,''
Astron.\ Astrophys.\  {\bf 401}, 835 (2003)
[arXiv:astro-ph/0302426].

\bibitem{Campanelli:2004pm}
  L.~Campanelli, A.~D.~Dolgov, M.~Giannotti and F.~L.~Villante,
  ``Faraday Rotation of the CMB Polarization and Primordial Magnetic Field Properties,''
  Astrophys.\ J.\  {\bf 616}, 1 (2004)
  [arXiv:astro-ph/0405420].

\bibitem{Kosowsky:2004zh}
A.~Kosowsky, T.~Kahniashvili, G.~Lavrelashvili and B.~Ratra,
``Faraday Rotation of the Cosmic Microwave Background Polarization by a Stochastic Magnetic Field,''
Phys.\ Rev.\ D {\bf 71}, 043006 (2005) [arXiv:astro-ph/0409767].

\bibitem{kmk08}
T.~Kahniashvili, Y.~Maravin and A.~Kosowsky,
``Primordial Magnetic Field Limits from WMAP Five-Year Data,''
Phys.\ Rev.\ D {\bf 80}, 023009 (2009)
[arXiv:0806.1876 [astro-ph]].

\bibitem{Cabella:2007br}
P.~Cabella, P.~Natoli and J.~Silk,
``Constraints on CPT Violation from WMAP Three-Year Polarization Data: A Wavelet Analysis,''
Phys.\ Rev.\  D {\bf 76}, 123014 (2007)
[arXiv:0705.0810 [astro-ph]].

\bibitem{Xia:2007qs}
J.~Q.~Xia, H.~Li, X.~l.~Wang and X.~m.~Zhang,
``Testing CPT symmetry with CMB measurements,''
arXiv:0710.3325 [hep-ph].

\bibitem{Feng:2006dp}
B.~Feng, M.~Li, J.~Q.~Xia, X.~Chen and X.~Zhang,
``Searching for CPT violation with WMAP and BOOMERANG,''
Phys.\ Rev.\ Lett.\  {\bf 96}, 221302 (2006)
[arXiv:astro-ph/0601095].

\bibitem{Xia:2012ck}
J.~-Q.~Xia,
``Cosmological CPT Violation and CMB Polarization Measurements,''
JCAP {\bf 1201}, 046 (2012)  [arXiv:1201.4457 [astro-ph.CO]].

\bibitem{Xia:2009ah}
J.~-Q.~Xia, H.~Li and X.~Zhang,
``Probing CPT Violation with CMB Polarization Measurements,''
Phys.\ Lett.\ B {\bf 687}, 129 (2010)  [arXiv:0908.1876 [astro-ph.CO]].

\bibitem{Li:2009rt}
M.~Li, Y.~-F.~Cai, X.~Wang and X.~Zhang,
``$CPT$ Violating Electrodynamics and Chern-Simons Modified Gravity,''
Phys.\ Lett.\ B {\bf 680}, 118 (2009)  [arXiv:0907.5159 [hep-ph]].

\bibitem{Li:2008tma}
M.~Li and X.~Zhang,
``Cosmological CPT-Violating Effect on CMB Polarization,''
Phys.\  Rev.\ D {\bf 78}, 103516 (2008)  [arXiv:0810.0403 [astro-ph]].

\bibitem{Gruppuso:2011ci}
A.~Gruppuso, P.~Natoli, N.~Mandolesi, A.~De Rosa, F.~Finelli and F.~Paci,
``WMAP 7 Year Constraints on CPT Violation from Large Angle CMB  Anisotropies,''
JCAP {\bf 1202}, 023 (2012) [arXiv:1107.5548 [astro-ph.CO]].

\bibitem{Lue:1998mq}
A.~Lue, L.~M.~Wang and M.~Kamionkowski,
``Cosmological Signature of New Parity-Violating Interactions,''
Phys.\ Rev.\ Lett.\  {\bf 83}, 1506 (1999)
[arXiv:astro-ph/9812088].

\bibitem{Xia:2008si}
J.~Q.~Xia, H.~Li, G.~B.~Zhao and X.~Zhang,
``Testing CPT Symmetry with CMB Measurements: Update After WMAP5,''
arXiv:0803.2350 [astro-ph].

\bibitem{Saito:2007kt}
S.~Saito, K.~Ichiki and A.~Taruya,
``Probing polarization states of primordial gravitational waves with CMB anisotropies,''
JCAP {\bf 0709}, 002 (2007)
[arXiv:0705.3701 [astro-ph]].

\bibitem{Scannapieco:1997mt}
E.~S.~Scannapieco and P.~G.~Ferreira,
``Polarization - temperature correlation from primordial magnetic field,''
Phys.\ Rev.\  D {\bf 56}, 7493 (1997)
[arXiv:astro-ph/9707115].

\bibitem{Scoccola:2004ke}
C.~Scoccola, D.~Harari and S.~Mollerach,
``B polarization of the CMB from Faraday rotation,''
Phys.\ Rev.\  D {\bf 70}, 063003 (2004)
[arXiv:astro-ph/0405396].

\bibitem{Demianski:2007fz}
M.~Demianski and A.~G.~Doroshkevich,
``Possible extensions of the standard cosmological model: anisotropy,
rotation, and magnetic field,''
Phys.\ Rev.\  D {\bf 75}, 123517 (2007)
[arXiv:astro-ph/0702381].

\bibitem{Kristiansen:2008tx}
J.~R.~Kristiansen and P.~G.~Ferreira,
``Constraining primordial magnetic fields with CMB polarization experiments,''
Phys.\ Rev.\ D {\bf 77}, 123004 (2008)
  [arXiv:0803.3210 [astro-ph]].


\bibitem{Carroll:1989vb}
S.~M.~Carroll, G.~B.~Field and R.~Jackiw,
``Limits on a Lorentz and parity violating modification of electrodynamics,''
Phys.\ Rev.\  D {\bf 41}, 1231 (1990).

\bibitem{Kostelecky:2007zz}
V.~A.~Kostelecky and M.~Mewes,
``Lorentz-violating electrodynamics and the cosmic microwave background,''
Phys.\ Rev.\ Lett.\  {\bf 99}, 011601 (2007)
[arXiv:astro-ph/0702379].

\bibitem{Cai:2009uc}
Y.~-F.~Cai, M.~Li and X.~Zhang,
``Testing the Lorentz and CPT Symmetry with CMB polarizations and a
non-relativistic Maxwell Theory,''
JCAP {\bf 1001}, 017 (2010)
[arXiv:0912.3317 [hep-ph]].

\bibitem{Ni:2007ar}
W.~-T.~Ni,
``From Equivalence Principles to Cosmology: Cosmic Polarization Rotation,
CMB Observation, Neutrino Number Asymmetry, Lorentz Invariance and CPT,''
Prog.\ Theor.\ Phys.\ Suppl.\  {\bf 172}, 49 (2008)
[arXiv:0712.4082 [astro-ph]].

\bibitem{Casana:2008ry}
R.~Casana, M.~M.~Ferreira, Jr. and J.~S.~Rodrigues,
``Lorentz-violating contributions of the Carroll-Field-Jackiw model to the CMB anisotropy,''
Phys.\ Rev.\ D {\bf 78}, 125013 (2008)
[arXiv:0810.0306 [hep-th]].

\bibitem{Caldwell:2011pu}
R.~R.~Caldwell, V.~Gluscevic and M.~Kamionkowski,
``Cross-Correlation of Cosmological Birefringence with CMB  Temperature,''
Phys.\ Rev.\ D {\bf 84}, 043504 (2011)  [arXiv:1104.1634 [astro-ph.CO]].

\bibitem{MosqueraCuesta:2011tz}
H.~J.~Mosquera Cuesta and G.~Lambiase,
``Nonlinear electrodynamics and CMB polarization,''
JCAP {\bf 1103}, 033 (2011)  [arXiv:1102.3092 [astro-ph.CO]].

\bibitem{Kamionkowski:2010rb}
M.~Kamionkowski and T.~Souradeep,
``The Odd-Parity CMB Bispectrum,''
Phys.\ Rev.\ D {\bf 83}, 027301 (2011)
[arXiv:1010.4304 [astro-ph.CO]].

\bibitem{Gluscevic:2010vv}
V.~Gluscevic and M.~Kamionkowski,
``Testing Parity-Violating Mechanisms with Cosmic Microwave Background Experiments,''
Phys.\ Rev.\ D {\bf 81}, 123529 (2010)  [arXiv:1002.1308 [astro-ph.CO]].

\bibitem{Mewes:2012sm}
M. Mewes, ``Optical-cavity tests of higher-order Lorentz violation,''
  Phys.\ Rev.\ D {\bf 85}, 116012 (2012)  [arXiv:1203.5331 [hep-ph]].

\bibitem{Gluscevic:2012me}
  V.~Gluscevic, D.~Hanson, M.~Kamionkowski and C.~M.~Hirata,
  ``First CMB Constraints on Direction-Dependent Cosmological Birefringence from WMAP-7,''
  Phys.\ Rev.\ D {\bf 86}, 103529 (2012)
  [arXiv:1206.5546 [astro-ph.CO]].

\bibitem{Ni:2009qm}
W.~-T.~Ni,
``Constraints on pseudoscalar-photon interaction from CMB polarization observation,''
arXiv:0910.4317 [gr-qc].

\bibitem{Miller:2009pt}
N.~J.~Miller, M.~Shimon and B.~G.~Keating,
``CMB Polarization Systematics Due to Beam Asymmetry:
Impact on   Cosmological Birefringence,''
Phys.\ Rev.\ D {\bf 79}, 103002 (2009)
[arXiv:0903.1116 [astro-ph.CO]].

\bibitem{Ni:2009gz}
W.~-T.~Ni,
``Cosmic Polarization Rotation, Cosmological Models, and the Detectability
of Primordial Gravitational Waves,''
Int.\ J.\ Mod.\ Phys.\ A {\bf 24}, 3493 (2009)
[arXiv:0903.0756 [astro-ph.CO]].

\bibitem{Lim:2004js}
E.~A.~Lim,
``Can we see Lorentz-violating vector fields in the CMB?,''
Phys.\ Rev.\  D {\bf 71}, 063504 (2005)
[arXiv:astro-ph/0407437].

\bibitem{Carroll:2004ai}
S.~M.~Carroll and E.~A.~Lim,
``Lorentz-violating vector fields slow the universe down,''
Phys.\ Rev.\  D {\bf 70}, 123525 (2004)
[arXiv:hep-th/0407149].

\bibitem{Alexander:2006mt}
S.~H.~S.~Alexander,
``Is cosmic parity violation responsible for the anomalies in the WMAP data?,''
Phys.\ Lett.\  B {\bf 660}, 444 (2008)
[arXiv:hep-th/0601034].

\bibitem{Satoh:2007gn}
M.~Satoh, S.~Kanno and J.~Soda,
``Circular polarization of primordial gravitational waves in string-inspired
inflationary cosmology,''
Phys.\ Rev.\  D {\bf 77}, 023526 (2008)
[arXiv:0706.3585 [astro-ph]].

\bibitem{komatsu}
 C.~L.~Bennett {\it et al.}  [WMAP Collaboration],
  ``Nine-Year Wilkinson Microwave Anisotropy Probe (WMAP) Observations: Final Maps and Results,''
  Astrophys.\ J.\ Suppl.\  {\bf 208}, 20 (2013)
  [arXiv:1212.5225 [astro-ph.CO]];

\bibitem{Hinshaw:2013}
G.~Hinshaw {\it et al.}  [WMAP Collaboration],
  ``Nine-Year Wilkinson Microwave Anisotropy Probe (WMAP) Observations:
   Cosmological Parameter Results,''
   Astrophys.\ J.\ Suppl.\  {\bf 208}, 19 (2013);
   [arXiv:1212.5226 [astro-ph.CO]].

\bibitem{Larson:2010gs}
D.~Larson, J.~Dunkley, G.~Hinshaw, E.~Komatsu, M.~R.~Nolta,
C.~L.~Bennett, B.~Gold and M.~Halpern {\it et al.},
``Seven-Year Wilkinson Microwave Anisotropy Probe (WMAP) Observations:
Power Spectra and WMAP-Derived Parameters,''
Astrophys.\ J.\ Suppl.\  {\bf 192}, 16 (2011)  [arXiv:1001.4635 [astro-ph.CO]].

\bibitem{gauge}
M. A. Berger,  "Introduction to magnetic helicity".
Plasma Physics and Controlled Fusion {\bf 41} 167 (1999).


\bibitem{durrer03}
R.~Durrer and C.~Caprini,
``Primordial Magnetic Fields and Causality,''
JCAP {\bf 0311}, 010 (2003) [arXiv:astro-ph/0305059].

\bibitem{ktr11}
T.~Kahniashvili, A.~G.~Tevzadze and B.~Ratra,
``Phase Transition Generated Cosmological Magnetic Field at Large Scales,''
Astrophys.\ J.\  {\bf 726}, 78 (2011) [arXiv:0907.0197 [astro-ph.CO]].

\bibitem{jedamzik98}
K.~Jedamzik, V.~Katalinic and A.~V.~Olinto,
``Damping of Cosmic Magnetic Fields,''
Phys.\ Rev.\  D {\bf 57}, 3264 (1998) [arXiv:astro-ph/9606080].

\bibitem{sub98b}
K.~Subramanian and J.~D.~Barrow,
``Microwave Background Signals from Tangled Magnetic Fields,''
Phys.\ Rev.\ Lett.\  {\bf 81}, 3575 (1998)
[arXiv:astro-ph/9803261].

\bibitem{dky98}
R.~Durrer, T.~Kahniashvili and A.~Yates,
``Microwave Background Anisotropies from Alfven waves,''
Phys.\ Rev.\  D {\bf 58}, 123004 (1998)
[arXiv:astro-ph/9807089].

\bibitem{Seshadri:2000ky}
T.~R.~Seshadri and K.~Subramanian,
``CMBR Polarization Signals from Tangled Magnetic Fields,''
Phys.\ Rev.\ Lett.\  {\bf 87}, 101301 (2001)
[arXiv:astro-ph/0012056].


\bibitem{Subramanian:2003sh}
K.~Subramanian, T.~R.~Seshadri and J.~D.~Barrow,
``Small-scale CMB polarization anisotropies due to tangled primordial
magnetic fields,''
Mon.\ Not.\ Roy.\ Astron.\ Soc.\  {\bf 344}, L31 (2003)
[arXiv:astro-ph/0303014].


\bibitem{hu97}
W.~Hu and M.~J.~White,
``CMB Anisotropies: Total Angular Momentum Method,''
Phys.\ Rev.\  D {\bf 56}, 596 (1997)
[arXiv:astro-ph/9702170].

\bibitem{Zaldarriaga:1995gi}
M.~Zaldarriaga and D.~D.~Harari,
``Analytic approach to the polarization of the cosmic microwave background in flat and open universes,''
Phys.\ Rev.\ D {\bf 52}, 3276 (1995)  [arXiv:astro-ph/9504085].

\bibitem{Cai:2012ci}
Z.~Cai and Y.~Zhang,
``Analytic Spectra of CMB Anisotropies and Polarization Generated by Scalar Perturbations in Synchronous Gauge,''
Class.\ Quant.\ Grav.\  {\bf 29}, 105009 (2012)
[arXiv:1204.6683 [astro-ph.CO]].

\bibitem{Feldman:1997qc}
G.~J.~Feldman and R.~D.~Cousins,
``A Unified approach to the classical statistical analysis of small signals,''
Phys.\ Rev.\ D {\bf 57}, 3873 (1998)
[arXiv:physics/9711021 [physics.data-an]].

\bibitem{current_limits} P.~A.~R.~Ade {\it et al.}  [Planck Collaboration],
  ``Planck 2013 results. XVI. Cosmological parameters,''
  arXiv:1303.5076 [astro-ph.CO].

\bibitem{Shaw:2010ea}
  J.~R.~Shaw and A.~Lewis,
  ``Constraining Primordial Magnetism,''
  Phys.\ Rev.\ D {\bf 86}, 043510 (2012)
  [arXiv:1006.4242 [astro-ph.CO]].

\bibitem{Yamazaki:2013hda}
  D.~G.~Yamazaki, K.~Ichiki and K.~Takahashi,
  ``Constraints on the multi-lognormal magnetic fields from the observations of the cosmic microwave background and the matter power spectrum,''
  Phys.\ Rev.\ D {\bf 88}, 103011 (2013)
  [arXiv:1311.2584 [astro-ph.CO]].

\bibitem{Kahniashvili:2012dy}
  T.~Kahniashvili, Y.~Maravin, A.~Natarajan, N.~Battaglia and A.~G.~Tevzadze,
  ``Constraining primordial magnetic fields through large scale structure,''
  Astrophys.\ J.\  {\bf 770}, 47 (2013)
  [arXiv:1211.2769 [astro-ph.CO]].

\bibitem{trivedi14}
P.~Trivedi, K.~Subramanian, and T.R.~Seshadri, ``Primordial Magnetic Field Limits from CMB Trispectrum
Scalar Modes and Planck Constraints,'' Phys.\ Rev.\ D {\bf 89}, 043523 (2014)
[arXiv:1312.5308 [astro-ph.CO]].

\bibitem{bonvin13}
C.~Bonvin, C.~Caprini, and R.~Durrer, ``Magnetic Fields from Inflation: The CMB Temperature Anisotropies,''
Phys.\ Rev.\ D {\bf 88}, 083515 (2013)
[arXiv:1308.3348 [astro-ph.CO]].

\bibitem{actpol} M.~Niemack et al., ``ACTPol: A Polarization-Sensitive Receiver or the
Atacama Cosmology Telescope,'' Proc.\ SPIE {\bf 7741}, 51 (2010) [arXiv:1006.5049 [astro-ph.IM]].

\bibitem{sptpol}  J.E.~Austermann et al., ``SPTPol: An Instrument for CMB Polarization Measurements with the
South Pole Telescope,'' Proc.\ SPIE {\bf 8452}, 84520E (2012) [arXiv:1210.4970 [astro-ph.IM]].

\bibitem{bicep2}  P.A.R.~Ade et al., ``BICEP2 I: Detection of B-Mode Polarization at Degree Angular Scales,"
Phys.\ Rev.\ Lett.\ {\bf 112}, 241101  (2014)
[arXiv:1403.3985 [astro-ph.CO]].

\bibitem{prism} P.~Andre et al., ``PRISM: A White Paper on the Ultimate Polarimetric Spectro-Imaging of the
Microwave and Far Infrared Sky'' [arXiv:1306.2259 [astro-ph.CO]].

\bibitem{gorbunov}
D.~S.~Gorbunov, V.~A.~Rubakov,
{\it Introduction To The Theory of The Early Universe:
Cosmological Perturbations and Inflationary Theory},
(World Scientific Publishing Company, 2011).




\end{thebibliography}
\end{document}